\documentstyle[amsfonts,aps,amsmath]{revtex}

\newcommand{\bei}{\begin{itemize}}
\newcommand{\eei}{\end{itemize}}
\newcommand{\bea}{\begin{eqnarray}}
\newcommand{\eea}{\end{eqnarray}}
\newcommand{\bequ}{\begin{equation}}
\newcommand{\eequ}{\end{equation}}

\begin{document}
\title{Exact stationary correlations for a far-from-equilibrium spin chain}
\author{B. Schmittmann$^1$ and F.~Schm\"{u}ser$^2$}
\address{$^1$Center for Stochastic Processes in Science and Engineering and 
\\
Department of Physics,\\
Virginia Tech, Blacksburg, Virginia 24061-0435, USA;\\
$^2$MPI f\"ur Physik komplexer Systeme, N\"othnitzer Str.~38,\\
01187 Dresden, Germany.}
\date{\today}
\maketitle

\begin{abstract}
A kinetic one-dimensional Ising model on a ring evolves according to a
generalization of Glauber rates, such that spins at even (odd) lattice sites
experience a temperature $T_{e}$ ($T_{o}$). Detailed balance is violated so
that the spin chain settles into a {\em non-equilibrium} stationary state,
characterized by multiple interactions of increasing range and spin order.
We derive the equations of motion for {\em arbitrary} correlation functions
and solve them to obtain an exact representation of the steady state. Two
nontrivial amplitudes reflect the sublattice symmetries; otherwise,
correlations decay exponentially, modulo the periodicity of the ring. In the
long chain limit, they factorize into products of two-point functions, in
precise analogy to the equilibrium Ising chain. The exact solution confirms
the expectation, based on simulations and renormalization group arguments,
that the long-time, long-distance behavior of this two-temperature model is
Ising-like, in spite of the apparent complexity of the stationary
distribution.
\end{abstract}




\section{Introduction}

Exact solutions of simple model systems play a key role in statistical
mechanics. They provide reliable quantitative information about regular as
well as singular behavior and serve as proving grounds for approximation
schemes. Moreover, they can set important pointers in areas where a general
theoretical framework is still lacking. A prime example is the study of
non-equilibrium stationary states (NESS). Similar to systems in thermal
equilibrium, NESS are characterized by time-independent macroscopic
observables; however, at present these can only be computed {\em with
explicit reference} to the imposed dynamics: there is, as yet, no equivalent
of Gibbs ensemble theory for far-from-equilibrium steady states. As a
consequence, most progress to date is made by studying specific models.

Since the Ising model \cite{Ising} is one of the most intimately known
interacting many-particle systems, many non-equilibrium models depart from
it by imposing an external force which drives the system out of equilibrium.
Examples include couplings to multiple heat baths \cite{TT-cons,SFTT,blote},
competing Glauber and Kawasaki dynamics \cite{comp-dyn}, or a
current-inducing global bias \cite{KLS}. In all of these models, reviewed in %
\cite{sz-rev}, the non-equilibrium perturbation violates the detailed
balance symmetry of the equilibrium dynamics. On large time and length
scales, one can probe how this affects {\em universal} behavior, or,
adopting a more microscopic but no less fundamental perspective, one can ask
how {\em non-universal} properties, such as the exact configurational
probabilities, are modified. One finds, generically, that non-equilibrium
forces have especially profound effects if ({\em i}) they couple to the bulk
rather than the boundaries \cite{T-grad,luc} of the system \cite{spohn}, and
({\em ii}) the dynamics satisfies a conservation law \cite{KettRG,KLSRG,UCT}%
. For example, Ising lattice gases with {\em conserved} particle number can
be driven into two distinct non-equilibrium universality classes, depending
on the symmetries of the external forces \cite{KLS,KettRG}. In contrast,
Ising-like systems with {\em non-conserved} dynamics remain in the
universality class of the equilibrium Ising model, even if the usual ${\Bbb Z%
}_{2}$-symmetry of the Ising model is broken \cite{cr-beh-noncons}. Even
then, however, the violation of detailed balance leads to fundamental
changes in the configurational probabilities which must be identified and
interpreted before we may hope to formalize our understanding of NESS. It is
here that exact solutions are expected to be most helpful.

Motivated by these considerations, we recently\thinspace \cite{fb}
investigated a very simple non-equilibrium Ising-like model, namely, a
one-dimensional {\em interacting} spin chain with spin-flip dynamics coupled
to {\em two} temperature baths. The rates are a simple but nontrivial
generalization of the familiar Glauber \cite{glauber} rates: spins at odd
(even) sites are coupled to a temperature $T_{o}$ ($T_{e}$). In two
dimensions, this model \cite{SFTT,blote} exhibits an order-disorder phase
transition which belongs to the Ising class, according to renormalization
group arguments \cite{cr-beh-noncons} and Monte Carlo simulations \cite%
{blote}. In one dimension, the two-spin correlation function can be
calculated exactly \cite{raczzia}. Its only singularity lies at $%
T_{o}=T_{e}=0$, so that the lower critical dimension is $d=1$. Seeking an
expression for the steady state, we solved the master equation
perturbatively, in an expansion in the temperature {\em difference} of the
two heat baths, up to and including second order terms \cite{fb}. To our
surprise, the full stationary distribution turned out to be rather complex:
At each order, additional spin operators appear, characterized by longer
spatial range interactions and higher-order spin products, and lower-order
coupling constants acquire corrections. So, at first order, one encounters
next-nearest neighbor pair interactions, while at second order, two new
terms appear: a next-next-nearest neighbor pair interaction, and a four-spin
interaction spanning four nearest-neighbor sites. All of these are allowed
by symmetry, and none allowed by symmetry are absent. Given this structure,
one can, at least in principle, extrapolate to higher orders in perturbation
theory.

Of course, these findings immediately raise an obvious question: how does
this relatively complicated stationary distribution generate long-wavelength
behavior in the Ising universality class? In this article, we pursue an
alternate route towards the answer:\ Instead of aiming for the stationary
state directly, we seek its representation in terms of correlation functions
- since the knowledge of all stationary correlation functions is equivalent
to knowing the steady state. Starting from the master equation, we derive a
hierarchy of equations of motion for correlation functions of arbitrary
numbers of spins. In $d=1$, this hierarchy is closed and soluble.
Remarkably, we find that - in contrast to the apparent complexity of the
stationary distribution itself - the correlation functions are very simple.
Apart from two nontrivial amplitudes which reflect the temperature
difference between the two sublattices, the structure of correlations is 
{\em completely analogous} to the equilibrium Ising model: specifically, for 
$N\rightarrow \infty $, arbitrary (even) $m$-point correlation functions
factorize into a product of $m/2$ two-point correlations. Of course, all
correlation functions involving an odd number of spins vanish by symmetry.

While some exact analytic results for steady-state distributions are
available, they are confined to three classes of systems: first,
one-dimensional lattice gas models, restricted to excluded volume
interactions, such as the asymmetric exclusion process and its relatives %
\cite{schuetz,priv,LS}, second, very special one-dimensional spin systems
whose master equations are solved by the Ising Boltzmann factor \cite%
{KLS,matt,schmju}, and third, interacting systems in one or two dimensions
on {\em very small} lattices so that the number of degrees of freedom
remains manageable \cite{small}. To the best of our knowledge, the work
presented here is amongst the first complete solutions for non-equilibrium
stationary states with nontrivial (nearest-neighbor) interactions and
arbitrary number of degrees of freedom.

The article is organized as follows. We first introduce our model and its
master equation. Next, we derive the equations of motion for arbitrary
correlation functions. Following a brief review of the solution \cite%
{raczzia} for the two-point correlation function, we show how four-point
correlations can be factorized into two-point correlations in the long chain
( $N\rightarrow \infty $ ) limit. We then postulate that {\em all}
correlation functions factorize in this manner, and show that this
factorization solves the equations of motion. In an appendix, we generalize
this solution to finite systems with periodic boundary conditions. We
conclude with some comments and open questions.

\section{The model}

\label{secmodel} Our model is defined on a one-dimensional ring, with an
even number $N$ of sites, and periodic boundary conditions. A spin variable, 
$\sigma _{i}=\pm 1$, denotes the value of the spin at site $i$, and
nearest-neighbor spins interact according to the usual Ising Hamiltonian, 
\begin{equation}
{\cal H}=-J\sum_{i}\sigma _{i}\sigma _{i+1}  \label{Ham}
\end{equation}
with an exchange coupling $J$. The dynamics is a non-equilibrium
generalization of the usual Glauber model \cite{glauber}: spins on even and
odd lattice sites experience{\em \ different} temperatures, $T_{e}$ and $%
T_{o}$. Specifically, a configuration $\{\sigma \}=\{\sigma _{1},\,\sigma
_{2},\dots ,\sigma _{N}\}$ evolves into a new configuration by flipping a
randomly selected spin $\sigma _{i}$ with a rate \cite{raczzia} 
\begin{equation}
w_{i}(\sigma _{i}\rightarrow -\sigma _{i})=1-\frac{\gamma _{i}}{2}\sigma
_{i}(\sigma _{i-1}+\sigma _{i+1})\;,  \label{rates}
\end{equation}
where 
\begin{equation}
\quad \gamma _{i}=\left\{ 
\begin{array}{l}
\;\gamma _{e}=\tanh (2J/k_{B}T_{e})\,,\quad \quad i\;\;{even} \\ 
\;\gamma _{o}=\tanh (2J/k_{B}T_{o})\,,\quad \quad i\;\;{odd}\quad .%
\end{array}
\right.  \label{gamma_i}
\end{equation}
Thus, the full time-dependent configurational probability $p(\{\sigma \};t)$
evolves according to a master equation, 
\begin{equation}
\partial _{t}p(\{\sigma \};t)=\sum_{i=1}^{N}\left[ -w_{i}(\sigma
_{i}\rightarrow -\sigma _{i})p\left( \{\sigma \};t\right) +w_{i}(-\sigma
_{i}\rightarrow \sigma _{i})p\left( \{\sigma ^{[i]}\};t\right) \right]
\label{me}
\end{equation}
where $\{\sigma ^{[i]}\}$ differs from $\{\sigma \}$ by a flip of the $i$-th
spin. A trivial time scale has been set to unity, and we use dimensionless
units for inverse temperature, i.e., $\beta _{e}\equiv J/(k_{B}\,T_{e})$,
etc.

Our goal in the following is to find a representation for the stationary
solution of Eqn~(\ref{me}), $q(\{\sigma \})\equiv \lim_{t\rightarrow \infty
}p(\{\sigma \};t)$. This limit is unique, since Eqn~(\ref{me}) is ergodic:
every configuration $\{\sigma \}$ can be reached in finite time from every
other configuration $\{\sigma ^{\prime }\}$ (unless $T_{e}=T_{o}=0$). For 
{\em equal} temperatures $T\equiv T_{e}=T_{o}$, the steady state is just the
(canonical) distribution for the Ising chain, 
\begin{equation}
q_{o}\left( \{\sigma \}\right) =\frac{1}{Z}\exp \left( -H/k_{B}T\right)
\quad .  \label{Ising}
\end{equation}

It is of course straightforward to compute arbitrary correlation functions
for the equilibrium Ising chain. Since the Ising model is invariant under a
global spin flip (${\Bbb Z}_{2}$-symmetry), only correlations of {\em even}
numbers of spins are nonzero, and are easily expressed in terms of the
parameter $\bar{\omega}=\tanh (J/k_{B}T)$. Of course, we may -- and always
will -- order the arguments of an $m$-point correlation function ($m$ even)
without loss of generality, such that $1\leq k_{1}<k_{2}<...<k_{m}\leq N$.
Then, in the $N\rightarrow \infty $ limit one finds easily: 
\begin{eqnarray}
\left\langle \sigma _{k_{1}}...\sigma _{k_{m}}\right\rangle ^{eq} &=&\bar{%
\omega}^{\left( k_{2}-k_{1}\right) +\left( k_{4}-k_{3}\right) +...+\left(
k_{m}-k_{m-1}\right) }  \nonumber \\
&=&\left\langle \sigma _{k_{1}}\sigma _{k_{2}}\right\rangle
^{eq}\left\langle \sigma _{k_{3}}\sigma _{k_{4}}\right\rangle
^{eq}...\left\langle \sigma _{k_{m-1}}\sigma _{k_{m}}\right\rangle
^{eq}\quad \;\text{ for }m\text{ even and }N\rightarrow \infty \;,\text{ }
\label{Ising-corr}
\end{eqnarray}
i.e., a general correlation function factorizes into a product of two-point
functions.

We now turn to the non-equilibrium model, characterized by {\em different
temperatures}, $T_{e}\neq T_{o}$. The associated stationary state violates
detailed balance \cite{fb} and differs from the Boltzmann distribution, Eqn~(%
\ref{Ising}). The degree to which detailed balance is violated can be
measured by the parameter $d\equiv (\gamma _{o}-\gamma _{e})/2$. Similar to
the Ising model, the stationary state is invariant under a global spin flip (%
${\Bbb Z}_{2}$). As a consequence, all stationary correlations of an odd
number of spins vanish identically, and again, only even correlations need
to be discussed. We also note the symmetry under translations by $n$ lattice
sites, ${\Bbb T}_{n}$, combined with $d\rightarrow -d$ if $n$ is odd: 
\begin{eqnarray}
q(\{\sigma \};d) &=&q(\{-\sigma \};d)  \label{sym} \\
q(\{\sigma \};d) &=&q({\Bbb T}_{n}\{\sigma \};(-1)^{n}d)  \nonumber
\end{eqnarray}%
While a direct exact solution of the master equation has proven elusive, a
perturbative calculation, in powers of $d$, shows \cite{fb} that the
stationary distribution for this non-equilibrium model is rather
complicated, with longer-range and higher-order spin operators appearing.
Specifically, writing $q\left( \{\sigma \}\right) \equiv \;\tilde{Z}%
^{-1}\exp \left[ V\left( \{\sigma \}\right) \right] $, we find that the
potential function $V\left( \{\sigma \}\right) $, to second order in $d$,
has the form 
\begin{equation}
V\left( \{\sigma \}\right) =\bar{\beta}\sum_{i}\sigma _{i}\sigma
_{i+1}+d\lambda \sum_{i}(-1)^{i}\,\sigma _{i}\sigma _{i+2}\,+2\left(
d\lambda \right) ^{2}\sum_{i}\left[ \sigma _{i}\sigma _{i+2}+\sigma
_{i}\sigma _{i+1}\sigma _{i+2}\sigma _{i+3}-\coth (2\bar{\beta})\left(
\sigma _{i}\sigma _{i+1}+\sigma _{i}\sigma _{i+3}\right) \right] \;,
\label{pot}
\end{equation}%
where $\tanh (2\bar{\beta})\equiv \left( \gamma _{e}+\gamma _{o}\right) /2$
and $\lambda =-\frac{1}{8}\sinh (4\,\bar{\beta})$. An analysis of the
structure of the perturbation series indicates that, at each order,
additional spin operators appear, consisting of spatially longer-range
interactions and higher-order spin products. It is therefore quite
remarkable that the correlation functions turn out to be very similar to
those of the Ising model, as we now proceed to show.

\section{Equations of motion for arbitrary correlation functions}

We begin by deriving the equations of motion for arbitrary time-dependent $m$%
-point spin correlation functions, $\left\langle \sigma _{k_{1}}\sigma
_{k_{2}}...\sigma _{k_{m}}\right\rangle _{t}$, starting from the master
equation: 
\[
\partial _{t}\left\langle \sigma _{k_{1}}\sigma _{k_{2}}...\sigma
_{k_{m}}\right\rangle _{t}=\sum_{i=1}^{N}\left\{ \sum_{\{\sigma \}}\sigma
_{k_{1}}\sigma _{k_{2}}...\sigma _{k_{m}}\left[ -w_{i}(\sigma
_{i}\rightarrow -\sigma _{i})p\left( \{\sigma \};t\right) +w_{i}(-\sigma
_{i}\rightarrow \sigma _{i})p\left( \{\sigma ^{\lbrack i]}\};t\right) \right]
\right\} 
\]
Here, the subscript $\left\langle \circ \right\rangle _{t}$ will be used to
distinguish time-dependent averages from their stationary limits, $%
\left\langle \circ \right\rangle \equiv \lim_{t\rightarrow \infty
}\left\langle \circ \right\rangle _{t}$. Due to the sum over all
configurations, the \{\} bracket on the right hand side obviously vanishes
for all sites $i$ which do not belong to the set $\{k_{1},k_{2},...,k_{m}\}$%
, and one finds easily that 
\[
\partial _{t}\left\langle \sigma _{k_{1}}\sigma _{k_{2}}...\sigma
_{k_{m}}\right\rangle _{t}=-2\sum_{i=1}^{m}\left\langle \sigma
_{k_{1}}\sigma _{k_{2}}...\sigma _{k_{m}}w_{k_{i}}(\sigma
_{k_{i}}\rightarrow -\sigma _{k_{i}})\right\rangle _{t}\quad . 
\]
Inserting Eqn~(\ref{rates}) for $w_{k_{i}}(\sigma _{k_{i}}\rightarrow
-\sigma _{k_{i}})$ and taking the infinite time limit, we obtain the
equations satisfied by the stationary correlation functions, $%
\lim_{t\rightarrow \infty }\left\langle \sigma _{k_{1}}\sigma
_{k_{2}}...\sigma _{k_{m}}\right\rangle _{t}\equiv \left\langle \sigma
_{k_{1}}\sigma _{k_{2}}...\sigma _{k_{m}}\right\rangle $: 
\begin{equation}
0=-2m\left\langle \sigma _{k_{1}}\sigma _{k_{2}}...\sigma
_{k_{m}}\right\rangle +\sum_{i=1}^{m}\gamma _{k_{i}}\left\langle \sigma
_{k_{1}}..\sigma _{k_{i-1}}(\sigma _{k_{i}+1}+\sigma _{k_{i}-1})\sigma
_{k_{i+1}}..\sigma _{k_{m}}\right\rangle  \label{corr's}
\end{equation}
As mentioned before, all stationary correlations of odd numbers of spins are
trivially zero, and we need to focus only on the case of even $m$. Moreover,
all stationary correlations are translationally invariant, modulo the
sublattice structure. As in the Glauber model \cite{gl-corr's}, the
hierarchy of correlation functions is closed: the equation for $m$-point
correlations does not involve any higher correlations, and it is homogeneous
provided there are no nearest-neighbor pairs among the arguments. Otherwise,
the $m$-point correlations couple to lower correlation functions, which
appear as inhomogeneities. In the following section, we motivate an ansatz
for $\left\langle \sigma _{k_{1}}\sigma _{k_{2}}...\sigma
_{k_{m}}\right\rangle $, and then show that it satisfies Eqn (\ref{corr's})
for any choice of $\{k_{1},k_{2},...,k_{m}\}$.

\section{Exact solutions for stationary correlations.\label{sec4}}

To establish several key relations, we briefly review the exact solution for
the two-point functions $\left\langle \sigma _{i}\sigma _{j}\right\rangle $ %
\cite{raczzia}. For this case, Eqn~(\ref{corr's}) reads 
\begin{equation}
0=-4\,\left\langle \sigma _{i}\sigma _{j}\right\rangle +\gamma
_{i}\left\langle (\sigma _{i+1}+\sigma _{i-1})\sigma _{j}\right\rangle
+\gamma _{j}\left\langle \sigma _{i}(\sigma _{j+1}+\sigma
_{j-1})\right\rangle \quad .  \label{2-pt-h}
\end{equation}%
For nearest-neighbor sites, e.g., $j=i+1$, this equation becomes
inhomogeneous: 
\begin{equation}
0=-4\left\langle \sigma _{i}\sigma _{i+1}\right\rangle +\gamma
_{i}\left\langle \sigma _{i-1}\sigma _{i+1}\right\rangle +\gamma
_{i+1}\left\langle \sigma _{i}\sigma _{i+2}\right\rangle +\gamma _{i}+\gamma
_{i+1}  \label{2-pt-i}
\end{equation}%
For simplicity, we restrict the discussion in this section to correlation
functions in the thermodynamic limit $N\rightarrow \infty $. Thus we may
label the lattice sites by the integers, $i\in {\Bbb Z}$. Correlations on
finite periodic chains and some details for the $N\rightarrow \infty $ limit
will be addressed in Appendix \ref{app1}. Following \cite{raczzia}, it is
easy to show that the solution is unique and takes the form (for $i<j$,
without loss of generality) 
\begin{equation}
\langle \sigma _{i}\,\sigma _{j}\rangle =\sqrt{A_{i}\,A_{j}}\,\omega ^{j-i}
\label{2-pt-sol}
\end{equation}%
with the spatial decay length controlled by the parameter 
\begin{equation}
\omega =\frac{1}{\sqrt{\gamma _{e}\gamma _{o}}}\left( 1-\sqrt{1-\gamma
_{e}\gamma _{o}}\right)   \label{omega}
\end{equation}%
The key difference to the equilibrium Ising model is the emergence of {\em %
two amplitudes}, to match the sublattice symmetries, namely, 
\begin{equation}
A_{i}\equiv \;\left\{ 
\begin{array}{l}
\;A_{e}=(\gamma _{e}+\gamma _{o})/(2\gamma _{o})\,,\quad \text{{if}}\quad
i\;\;\text{even} \\ 
\;A_{o}=(\gamma _{e}+\gamma _{o})/(2\gamma _{e})\,,\quad \text{{if}}\quad
i\;\;\text{{odd}}%
\end{array}%
\right.   \label{ampl}
\end{equation}%
For later reference, we note that these relations imply a further identity
for {\em products} of pair correlations, namely, for any $i<j<j+1<l$, 
\begin{equation}
0=\left( \gamma _{j}+\gamma _{j+1}\right) \;\langle \sigma _{i}\,\sigma
_{l}\rangle -\gamma _{j}\;\langle \sigma _{i}\,\sigma _{j+1}\rangle
\;\langle \sigma _{j+1}\,\sigma _{l}\rangle -\gamma _{j+1}\;\langle \sigma
_{i}\,\sigma _{j}\rangle \;\langle \sigma _{j}\,\sigma _{l}\rangle \;.
\label{insert}
\end{equation}%
This relation will be needed in the discussion of four-point functions.

It is quite remarkable that the three quantities $\omega $, $A_{e}$, and $%
A_{o}$ also determine all higher correlation functions, through a structure
which is almost perfectly analogous to the equilibrium case. Deferring the
case of finite $N$ to Appendix \ref{app1}, we consider only an infinite
chain here. Focusing on even correlations, we assert that arbitrary $m$%
-point correlation functions, with $k_{1}<k_{2}<...<k_{m}$ and $m$ even, are
given by 
\begin{equation}
\left\langle \sigma _{k_{1}}...\sigma _{k_{m}}\right\rangle =\sqrt{%
A_{k_{1}}A_{k_{2}}...A_{k_{m}}}\;\;\omega ^{\left( k_{2}-k_{1}\right)
+\left( k_{4}-k_{3}\right) +...+\left( k_{m}-k_{m-1}\right) }=\left\langle
\sigma _{k_{1}}\sigma _{k_{2}}\right\rangle \left\langle \sigma
_{k_{3}}\sigma _{k_{4}}\right\rangle ...\left\langle \sigma _{k_{m-1}}\sigma
_{k_{m}}\right\rangle \quad .  \label{fact}
\end{equation}
i.e., higher correlation functions factorize into two-spin correlations as
in the equilibrium case, Eqn~(\ref{Ising-corr}). However, two features
distinguish these correlations from their equilibrium counterparts. First,
the spatial dependence is controlled by a different parameter, $\omega $. In
analogy to $\bar{\omega}$ for the equilibrium system, $\omega $ defines an 
{\em effective temperature} for the non-equilibrium system, via $\omega
\equiv \tanh (J/k_{B}T_{{\rm eff}})$. $T_{{\rm eff}}$ diverges with $T_{o}$
or $T_{e}$ but vanishes only if both $T_{o}$ {\em and} $T_{e}$ go to zero.
As a result, the correlation length $\xi $, defined via $\ln \omega \equiv
-\xi ^{-1}$, diverges only if {\em both} temperatures vanish. Second, and
more importantly, we note the appearance of the even/odd amplitudes $%
A_{e},\;A_{o}$, reflecting the sublattice identities of the two spins. These
amplitudes carry the primary information about the non-equilibrium nature of
our dynamics.

Before turning to a general proof of Eqn~(\ref{fact}), it is instructive to
confirm it explicitly for the four-point functions. In this case,
factorization implies 
\begin{equation}
\left\langle \sigma _{i}\sigma _{j}\sigma _{k}\sigma _{l}\right\rangle
=\left\langle \sigma _{i}\sigma _{j}\right\rangle \left\langle \sigma
_{k}\sigma _{l}\right\rangle \quad ,  \label{4-pt-fact}
\end{equation}
for $i<j<k<l$. When inserting this ansatz into the right hand side of Eqn~(%
\ref{corr's}), we need to distinguish whether $j$ and $k$ are
nearest-neighbor sites or not. If $k>j+1$, we obtain 
\begin{eqnarray}
&&-8\left\langle \sigma _{i}\sigma _{j}\sigma _{k}\sigma _{l}\right\rangle
+\gamma _{i}\left\langle (\sigma _{i+1}+\sigma _{i-1})\sigma _{j}\sigma
_{k}\sigma _{l}\right\rangle +\gamma _{j}\left\langle \sigma _{i}(\sigma
_{j+1}+\sigma _{j-1})\sigma _{k}\sigma _{l}\right\rangle  \nonumber \\
&&+\gamma _{k}\left\langle \sigma _{i}\sigma _{j}(\sigma _{k+1}+\sigma
_{k-1})\sigma _{l}\right\rangle +\gamma _{l}\left\langle \sigma _{i}\sigma
_{j}\sigma _{k}(\sigma _{l+1}+\sigma _{l-1})\right\rangle  \nonumber \\
&=&\left[ -4\left\langle \sigma _{i}\sigma _{j}\right\rangle +\gamma
_{i}\left\langle (\sigma _{i+1}+\sigma _{i-1})\sigma _{j}\right\rangle
+\gamma _{j}\left\langle \sigma _{i}(\sigma _{j+1}+\sigma
_{j-1})\right\rangle \right] \left\langle \sigma _{k}\sigma _{l}\right\rangle
\eqnum{17a} \\
&&+\left[ -4\left\langle \sigma _{k}\sigma _{l}\right\rangle +\gamma
_{k}\left\langle (\sigma _{k+1}+\sigma _{k-1})\sigma _{l}\right\rangle
+\gamma _{l}\left\langle \sigma _{j}\sigma _{k}(\sigma _{l+1}+\sigma
_{l-1})\right\rangle \right] \left\langle \sigma _{i}\sigma _{j}\right\rangle
\nonumber \\
&=&0\quad ,  \nonumber
\end{eqnarray}
since the expressions in the square brackets vanish for the stationary
two-point functions, by virtue of Eqns~(\ref{2-pt-h}) or (\ref{2-pt-i}),
depending on whether ($i,j$) or ($k,l$) are nearest neighbors or not. If $%
k=j+1$, we need to consider a slightly different equation, namely 
\begin{eqnarray}
&&-8\left\langle \sigma _{i}\sigma _{j}\sigma _{j+1}\sigma _{l}\right\rangle
+\gamma _{i}\left\langle (\sigma _{i+1}+\sigma _{i-1})\sigma _{j}\sigma
_{j+1}\sigma _{l}\right\rangle +\gamma _{j}\left\langle \sigma _{i}(\sigma
_{j+1}+\sigma _{j-1})\sigma _{j+1}\sigma _{l}\right\rangle  \nonumber \\
&&+\gamma _{j+1}\left\langle \sigma _{i}\sigma _{j}(\sigma _{j+2}+\sigma
_{j})\sigma _{l}\right\rangle +\gamma _{l}\left\langle \sigma _{i}\sigma
_{j}\sigma _{j+1}(\sigma _{l+1}+\sigma _{l-1})\right\rangle  \nonumber \\
&=&\left[ -4\left\langle \sigma _{i}\sigma _{j}\right\rangle +\gamma
_{i}\left\langle (\sigma _{i+1}+\sigma _{i-1})\sigma _{j}\right\rangle
+\gamma _{j}\left\langle \sigma _{i}\sigma _{j-1}\right\rangle \right]
\left\langle \sigma _{j+1}\sigma _{l}\right\rangle +\gamma _{j}\left\langle
\sigma _{i}\sigma _{l}\right\rangle  \nonumber \\
&&+\left\langle \sigma _{i}\sigma _{j}\right\rangle \left[ -4\left\langle
\sigma _{j+1}\sigma _{l}\right\rangle +\gamma _{j+1}\left\langle \sigma
_{j+2}\sigma _{l}\right\rangle +\gamma _{l}\left\langle \sigma _{j+1}(\sigma
_{l+1}+\sigma _{l-1})\right\rangle \right] +\gamma _{j+1}\left\langle \sigma
_{i}\sigma _{l}\right\rangle  \eqnum{17b} \\
&=&\gamma _{j}\left\langle \sigma _{i}\sigma _{l}\right\rangle +\gamma
_{j+1}\left\langle \sigma _{i}\sigma _{l}\right\rangle -\gamma
_{j}\left\langle \sigma _{i}\sigma _{j+1}\right\rangle \left\langle \sigma
_{j+1}\sigma _{l}\right\rangle -\gamma _{j+1}\left\langle \sigma _{i}\sigma
_{j}\right\rangle \left\langle \sigma _{j}\sigma _{l}\right\rangle  \nonumber
\\
&=&0\quad .  \nonumber
\end{eqnarray}
The last equality follows from Eqn~(\ref{insert}).

The proof is completed by induction. Assuming that the factorization has
been proven for $(m-2)$-point functions, it is sufficient to show that the
ansatz 
\begin{equation}
\left\langle \sigma _{k_{1}}...\sigma _{k_{m}}\right\rangle =\left\langle
\sigma _{k_{1}}...\sigma _{k_{m-2}}\right\rangle \left\langle \sigma
_{k_{m-1}}\sigma _{k_{m}}\right\rangle   \label{red}
\end{equation}%
solves the equations of motion for the $m$-point functions. Again, we need
to distinguish whether $k_{m-2}$ and $k_{m-1}$ are nearest neighbors or not.
If they are separated by more than one lattice spacing, i.e., if $%
k_{m-1}>k_{m-2}+1$, we have 
\begin{eqnarray}
&&-2m\left\langle \sigma _{k_{1}}\sigma _{k_{2}}...\sigma
_{k_{m}}\right\rangle +\sum_{i=1}^{m}\gamma _{k_{i}}\left\langle \sigma
_{k_{1}}..\sigma _{k_{i-1}}(\sigma _{k_{i}+1}+\sigma _{k_{i}-1})\sigma
_{k_{i+1}}..\sigma _{k_{m}}\right\rangle   \nonumber \\
&=&\left\langle \sigma _{k_{m-1}}\sigma _{k_{m}}\right\rangle \left[
-2\left( m-2\right) \left\langle \sigma _{k_{1}}...\sigma
_{k_{m-2}}\right\rangle +\sum_{i=1}^{m-2}\gamma _{k_{i}}\left\langle \sigma
_{k_{1}}..\sigma _{k_{i-1}}(\sigma _{k_{i}+1}+\sigma _{k_{i}-1})\sigma
_{k_{i+1}}..\sigma _{k_{m-2}}\right\rangle \right]   \nonumber \\
&&+\left\langle \sigma _{k_{1}}...\sigma _{k_{m-2}}\right\rangle \left[
-2\left\langle \sigma _{k_{m-1}}\sigma _{k_{m}}\right\rangle +\gamma
_{k_{m-1}}\left\langle (\sigma _{k_{m-1}+1}+\sigma _{k_{m-1}-1})\sigma
_{k_{m}}\right\rangle +\gamma _{k_{m}}\left\langle \sigma _{k_{m-1}}(\sigma
_{k_{m}+1}+\sigma _{k_{m}-1})\right\rangle \right]   \eqnum{18a} \\
&=&0  \nonumber
\end{eqnarray}%
Here, we have used the fact that both square brackets vanish, since they
enclose the equations of motion for the two- and $(m-2)$-point functions,
respectively. Next, we consider the case $k_{m-1}=k_{m-2}+1$. Again, we add
and subtract the terms missing from the full equations of motion of the
lower correlations functions: 
\begin{eqnarray}
&&-2m\left\langle \sigma _{k_{1}}\sigma _{k_{2}}...\sigma
_{k_{m}}\right\rangle +\sum_{i=1}^{m}\gamma _{k_{i}}\left\langle \sigma
_{k_{1}}..\sigma _{k_{i-1}}(\sigma _{k_{i}+1}+\sigma _{k_{i}-1})\sigma
_{k_{i+1}}..\sigma _{k_{m}}\right\rangle   \nonumber \\
&=&-\gamma _{k_{m-2}}\left\langle \sigma _{k_{1}}..\sigma _{k_{m-3}}\sigma
_{k_{m-2}+1}\right\rangle \left\langle \sigma _{k_{m-1}}\sigma
_{k_{m}}\right\rangle +\gamma _{k_{m-2}}\left\langle \sigma _{k_{1}}..\sigma
_{k_{m-3}}\sigma _{k_{m}}\right\rangle   \nonumber \\
&&-\gamma _{k_{m-1}}\left\langle \sigma _{k_{1}}..\sigma
_{k_{m-2}}\right\rangle \left\langle \sigma _{k_{m-1}-1}\sigma
_{k_{m}}\right\rangle +\gamma _{k_{m-1}}\left\langle \sigma _{k_{1}}..\sigma
_{k_{m-3}}\sigma _{k_{m}}\right\rangle   \nonumber \\
&=&\left\langle \sigma _{k_{1}}..\sigma _{k_{m-4}}\right\rangle \left\{
\left( \gamma _{k_{m-2}}+\gamma _{k_{m-1}}\right) \left\langle \sigma
_{k_{m-3}}\sigma _{k_{m}}\right\rangle \right.   \eqnum{18b} \\
&&\left. -\gamma _{k_{m-2}}\left\langle \sigma _{k_{m-3}}\sigma
_{k_{m-2}+1}\right\rangle \left\langle \sigma _{k_{m-1}}\sigma
_{k_{m}}\right\rangle -\gamma _{k_{m-1}}\left\langle \sigma _{k_{m-3}}\sigma
_{k_{m-2}}\right\rangle \left\langle \sigma _{k_{m-1}-1}\sigma
_{k_{m}}\right\rangle \right\}   \nonumber \\
&=&0\quad .  \nonumber
\end{eqnarray}%
To obtain the last two identities, we have factored out an $(m-4)$-point
correlation, and used Eqn~(\ref{insert}). This concludes the proof. We
emphasize again that the condition $N\gg 1$ has been imposed for reasons of
simplicity alone: expressions for arbitrary correlation functions on {\em %
finite} rings, while somewhat more cumbersome, are easily derived and also
follow the Ising pattern, as we will show in Appendix \ref{app1}.

Finally, we discuss one special case of the two temperature Glauber dynamics
where the two--point correlations do not have the form of Eqn (\ref{2-pt-sol}%
). This is the case, if one of the two parameters $\gamma_e, \, \gamma_o $
is zero, the other finite, i.~e.~the temperature of one sublattice is
infinite. Without loss of generality we assume in the following $\gamma_o = 0
$. One can then easily show directly from the equation of motion (\ref%
{2-pt-h}) that the two-point function reads 
\begin{equation}
\langle \sigma_i \, \sigma_j \rangle \equiv \;\left\{ 
\begin{array}{l}
\; \gamma_e / 4 \,,\quad \text{{if}}\quad j-i=1 \\ 
\; \gamma_e^2 / 8 \,,\quad \text{{if}}\quad j-i=2 \; \text{and}\; i \; \text{%
even} \\ 
\; \, 0\, , \quad \quad \text{otherwise}%
\end{array}
\right.  \label{diff2pt}
\end{equation}
Hence, spins are only correlated over a distance of at most two lattice
sites. Turning to the higher correlations, we observe that Eqn (\ref{insert}%
) also holds for these parameter values. Therefore, $m$-point correlations ($%
m$ even) again factorize into two-point correlations.

\section{Conclusions}

To summarize, we have found an exact solution for all stationary correlation
functions of a one-dimensional non-equilibrium Ising spin chain with an even
number $N$ of sites. The system is globally coupled to two temperature
baths: spins on odd (even) lattice sites experience a temperature $T_{o}$ ($%
T_{e}$) and flip according to a generalization of the familiar Glauber
rates. The presence of two different temperatures violates detailed balance
and maintains a nontrivial non-equilibrium stationary state. The complete
set of correlation functions provides us with the full, exact solution for
this steady state. This allows us to reconcile two potentially contradictory
earlier findings: while simulations and renormalization group arguments
indicate that the long-wavelength behavior of non-conserved two-temperature
models should be Ising-like, a perturbative analysis of the stationary
distribution $q\left( \{\sigma \}\right) $ showed that a zoo of {\em %
long-range and multi-spin} interactions are present in $\ln q\left( \{\sigma
\}\right) $. Here, we have shown that the exact correlation functions for
this model are remarkably Ising-like: they decay exponentially with a
characteristic correlation length which diverges only if both both $T_{o}$
and $T_{e}$ vanish. Also, in the $N\rightarrow \infty $ limit, arbitrary
(even) $m$-point correlations factorize into products of two-point
correlations, following exactly the same scheme in both the Ising and the
two-temperature chain. The only key difference is the appearance of two
nontrivial amplitudes, $A_{o}$ and $A_{e}$, which reflect the sublattice
symmetry: for each spin on an odd (even) site, the correlation function
carries a factor of $\sqrt{A_{o}}$ ($\sqrt{A_{e}}$). Of course, stationary
correlations of odd numbers of spins vanish identically, due to symmetry. In
short, the correlations of the model are entirely consistent with its
Ising-like long-distance properties.

It is remarkable - and not at all immediately obvious - that these two
amplitudes should be the only remnants of the zoo of interactions in $%
q\left( \{\sigma \}\right) $. There is, however, a very simple and elegant
representation \cite{hjh} of the stationary state in terms of an extended
Ising model, consisting of $2N$ spins on a comb-like $d=1$ lattice with
Hamiltonian ${\cal H}_{aux}=-J\sum_{i}s_{i}s_{i+1}-J_{o}\sum_{\text{odd }%
i}s_{i}\sigma _{i}-J_{e}\sum_{\text{even }i}s_{i}\sigma _{i}$. Here, the
spins $\{s_{i}\}$ form an auxiliary set which must be traced out in order to
obtain stationary observables associated with the original variables $%
\{\sigma _{i}\}$. If the interactions $J$, $J_{o}$ and $J_{e}$ are tuned
appropriately {\em in the complex plane}, the $\sigma $-correlations of $%
{\cal H}_{aux}$ are identical to those of the two-temperature model.
Further, the exact stationary state of the two-temperature model follows as $%
q\left( \{\sigma \}\right) \propto \text{Tr}_{\{s_{i}\}} \exp [-{\cal H}%
_{aux}]$. Details and generalizations will be presented elsewhere \cite{hsz}.

It would of course be interesting to investigate other non-equilibrium
versions of Glauber dynamics. Will the correlations still have a rather
simple structure if a spin chain is coupled, in a translation-invariant
manner, to $n>2$ different temperatures? Will they still factorize, provided
a sufficiently large number of amplitudes is introduced? Clearly, we still
need to cover lots of ground before even such simple non-equilibrium systems
are fully understood.

{\bf Acknowledgements} We thank H.J. Hilhorst, R.K.P. Zia, J. Slawny, U.C. T%
\"{a}uber, W.~Just, E. Ben-Naim and M. Hastings for fruitful discussions.
Financial support from the NSF through the Division of Materials Research
and from the DFG through contract no. SCHM 1537 is gratefully acknowledged. %
\appendix

\section{Stationary correlations in finite chains}

\label{app1} In this Appendix, we address the correlation functions for our
model on a {\em finite} periodic chain (a ring) of $N$ sites, with $%
i=1,2,...,N$. To set the scene, we first review the equilibrium case, with
uniform temperature $T$. The correlation functions of the Ising model on a
ring are well known. Using an {\em ordered} set of arguments, $1\leq
k_{1}<k_{2}<...<k_{m}\leq N$ without loss of generality, one finds: 
\begin{eqnarray}
\left\langle \sigma _{k_{1}}...\sigma _{k_{m}}\right\rangle ^{eq} &=&0\text{
for odd }m\text{ }\;,  \nonumber \\
\left\langle \sigma _{k_{1}}...\sigma _{k_{m}}\right\rangle ^{eq} &=&\frac{1%
}{1+\bar{\omega}^{N}}\left\{ \bar{\omega}^{\left( k_{2}-k_{1}\right) +\left(
k_{4}-k_{3}\right) +...+\left( k_{m}-k_{m-1}\right) }+\bar{\omega}^{N-\left(
k_{2}-k_{1}\right) -\left( k_{4}-k_{3}\right) -...-\left(
k_{m}-k_{m-1}\right) }\right\} \text{ for even }m\quad .  \label{Isiper}
\end{eqnarray}%
with $\bar{\omega}=\tanh (J/k_{B}T)<1$. The periodicity of the ring implies
that $\left\langle \sigma _{k_{1}}...\sigma _{k_{m}}\right\rangle ^{eq}$ is
invariant under $\bar{\omega}^{n}\rightarrow \bar{\omega}^{N-n}$ for any
integer $0\leq n\leq N$. For $N\rightarrow \infty $, Eqn (\ref{Isiper})
obviously reduces to Eqn (\ref{Ising-corr}).

Turning to the non-equilibrium case with two temperatures, we first assert
that the stationary two-point correlation function, with $1\leq i<j\leq N$,
is given by 
\begin{equation}
\langle \sigma _{i}\,\sigma _{j}\rangle =\frac{1}{Z(\omega )}\sqrt{%
A_{i}\,A_{j}}\;\left[ \omega ^{j-i}+\omega ^{N-(j-i)}\right] \;.
\label{2-pt-solp}
\end{equation}
where the parameter $\omega $ and the amplitudes $A_{i}\in \{A_{e},\,A_{o}\}$
are simply those of Eqns (\ref{omega}) and (\ref{ampl}), and the
normalization factor $Z(\omega )$ is given by $Z(\omega )=1+\omega ^{N}$.
Clearly, $\langle \sigma _{i}\,\sigma _{j}\rangle $ is periodic with period $%
N$. For easy reference, we fill in some details here which are also relevant
to the $N\rightarrow \infty $ case.

Eqn (\ref{2-pt-h}) is essentially a second order difference equation in the
variable $j-i$. To find a unique solution, we need two boundary conditions:
one of these is Eqn (\ref{2-pt-i}) for $j-i=1$ and the other is the
requirement that $\langle \sigma _{i}\,\sigma _{j}\rangle $ be periodic with
period $N$ (or, for $N\rightarrow \infty $, that $\langle \sigma
_{i}\,\sigma _{j}\rangle $ vanish for large separations). We therefore
expect a superposition of two linearly independent solutions whose
coefficients can then be determined. Away from the boundaries, the ansatz $%
\left\langle \sigma _{i}\sigma _{j}\right\rangle =\sqrt{A_{i}A_{j}}\omega
^{j-i}$ reduces Eqn (\ref{2-pt-h}) to three relations, for the combinations $%
i,j$ odd/odd, even/even, and odd/even respectively: 
\begin{eqnarray*}
0 &=&-4A_{o}+2\gamma _{o}\sqrt{A_{e}\,A_{o}}\left( \omega +\omega
^{-1}\right) \text{ } \\
0 &=&-4A_{e}+2\gamma _{e}\sqrt{A_{e}\,A_{o}}\left( \omega +\omega
^{-1}\right)  \\
0 &=&-4\sqrt{A_{e}\,A_{o}}+\left( \gamma _{o}A_{e}+\gamma _{e}A_{o}\right)
\left( \omega +\omega ^{-1}\right) 
\end{eqnarray*}%
We note that the third equation is simply a linear combination of the first
and second, expressing the fact that there are only {\em two} amplitudes,
and not three, as one might have assumed (cf. \cite{raczzia}) based on the
three types (odd/odd, even/even, odd/even) of pair correlations. Moreover,
the symmetry under $\omega \leftrightarrow \omega ^{-1}$ implies that, for
each solution $\omega $, $\omega ^{-1}$ \ is also a solution. Proceeding to
solve the system, we find $\gamma _{e}A_{o}=\gamma _{o}A_{e}$ and 
\[
\omega +\omega ^{-1}=\frac{2}{\sqrt{\gamma _{o}\gamma _{e}}}\Rightarrow
\omega _{\pm }=\frac{1}{\sqrt{\gamma _{e}\gamma _{o}}}\left( 1\pm \sqrt{%
1-\gamma _{e}\gamma _{o}}\right) 
\]%
with $\omega _{+}=1/\omega _{-}$. The two roots $\omega _{\pm }$ of this
quadratic equation provide us with the two anticipated linearly independent
solutions. However, to satisfy the inhomogeneous Eqn (\ref{2-pt-i}) for both
finite and infinite $N$, we have to take a convex combination of the two
solutions 
\begin{equation}
\left\langle \sigma _{i}\sigma _{j}\right\rangle =\alpha \sqrt{A_{i}A_{j}}%
\omega _{-}^{j-i}+\beta \sqrt{A_{i}A_{j}}\omega _{+}^{j-i}\quad \quad \text{%
with}\;\;\alpha +\beta =1\quad ,  \label{anscorr}
\end{equation}%
which then leads to $\gamma _{e}\,A_{o}+\gamma _{o}\,A_{e}=\gamma
_{e}+\gamma _{o}$. This relation, together with the previous identity $%
\gamma _{e}A_{o}=\gamma _{o}A_{e}$, determines the values of the two
amplitudes $A_{e}$ and $A_{o}$. Demanding periodicity for the ansatz (\ref%
{anscorr}) results in a second identity for the two integration constants $%
\alpha ,\,\beta $, namely $\beta =\alpha \omega _{-}^{N}$. Comparing the
result with our assertion, Eqn (\ref{2-pt-solp}), we identify $\omega \equiv
\omega _{-}$ and $Z(\omega )\equiv 1+\omega _{-}^{N}$. Since $\omega _{-}<1$%
, the $N\rightarrow \infty $ limit is also obvious.

To treat the general case, we set up some notation. For any even $m\geq 2$,
we define two auxiliary functions which depend on an ordered set of
arguments $1\leq k_{1}<k_{2}<...<k_{m}\leq N$, 
\begin{eqnarray*}
S_{m}^{(f)}(k_{1},k_{2},...k_{m}) &\equiv &\sqrt{A_{k_{1}}%
\,A_{k_{2}}...A_{k_{m}}}\omega ^{\left( k_{2}-k_{1}\right) +\left(
k_{4}-k_{3}\right) +...+\left( k_{m}-k_{m-1}\right)
}=S_{2}^{(f)}(k_{1},k_{2})S_{2}^{(f)}(k_{3},k_{4})...S_{2}^{(f)}(k_{m-1},k_{m})
\\
S_{m}^{(b)}(k_{1},k_{2},...k_{m}) &\equiv &\sqrt{A_{k_{1}}%
\,A_{k_{2}}...A_{k_{m}}}\omega ^{-\left( k_{2}-k_{1}\right) -\left(
k_{4}-k_{3}\right) -...-\left( k_{m}-k_{m-1}\right)
}=S_{2}^{(b)}(k_{1},k_{2})S_{2}^{(b)}(k_{3},k_{4})...S_{2}^{(b)}(k_{m-1},k_{m})
\end{eqnarray*}%
Thus, we rewrite Eqn (\ref{2-pt-solp}) in the form 
\[
\langle \sigma _{i}\,\sigma _{j}\rangle =\frac{1}{Z(\omega )}\;\left[
S_{2}^{(f)}(i,j)+\omega ^{N}S_{2}^{(b)}(i,j)\right] \quad .
\]%
and recall that {\em both} $S_{2}^{(f)}(i,j)\ $ {\em and} $S_{2}^{(b)}(i,j)$
satisfy the equations of motion for the ${\em infinite}$ chain, i.e.,  (\ref%
{2-pt-h}), (\ref{2-pt-i}), and (\ref{insert}). 

We now proceed as follows:\ First, we argue that both $%
S_{m}^{(f)}(k_{1},k_{2},...k_{m})$ and $S_{m}^{(b)}(k_{1},k_{2},...k_{m})$
solve the homogeneous and inhomogeneous equations of motion (\ref{corr's})
for $m$-point functions on the infinite chain. For $%
S_{m}^{(f)}(k_{1},k_{2},...k_{m})=\lim_{N\rightarrow \infty }\left\langle
\sigma _{k_{1}}...\sigma _{k_{m}}\right\rangle $, this is immediately
obvious. For $S_{m}^{(b)}(k_{1},k_{2},...k_{m})$, \ we simply need to
retrace the inductive proof from Section 4: Since the basic relations (\ref%
{2-pt-h}), (\ref{2-pt-i}), and (\ref{insert}) all hold for $S_{2}^{(b)}$, we
may replace each $\left\langle \sigma _{k_{1}}...\sigma
_{k_{m}}\right\rangle $ in Eqns (17-18) by the appropriate $S_{m}^{(b)}$
without violating any of the equalities. Since some of the equations of
motion (\ref{corr's}) for the $m$-point correlations are inhomogeneous, only
a convex combination $\alpha S_{m}^{(f)}(k_{1},k_{2},...k_{m})+\beta
S_{m}^{(b)}(k_{1},k_{2},...k_{m})$ with $\alpha +\beta =1$ is a solution.

Finally, we have to to satisfy the last remaining property of the full
correlation function, namely, periodicity. Given an ordered set of arguments 
$1\leq k_{1}<k_{2}<...<k_{m}\leq N$, we assert that a general correlation
function of $m\geq 2$ spins takes the form 
\begin{eqnarray}
\left\langle \sigma _{k_{1}}...\sigma _{k_{m}}\right\rangle &=&0\text{ for
odd }m\text{ }\;,  \nonumber \\
\left\langle \sigma _{k_{1}}...\sigma _{k_{m}}\right\rangle &=&\frac{\sqrt{%
A_{k_{1}}\,A_{k_{2}}...A_{k_{m}}}}{Z(\omega )}\left\{ \omega ^{\left(
k_{2}-k_{1}\right) +\left( k_{4}-k_{3}\right) +...+\left(
k_{m}-k_{m-1}\right) }+\omega ^{N-\left[ \left( k_{2}-k_{1}\right) +\left(
k_{4}-k_{3}\right) +...+\left( k_{m}-k_{m-1}\right) \right] }\right\}
\label{m-pt} \\
&=&\frac{1}{Z(\omega )}\left\{ S_{m}^{(f)}(k_{1},k_{2},...k_{m})+\omega
^{N}S_{m}^{(b)}(k_{1},k_{2},...k_{m})\right\} \;\text{for even }m  \nonumber
\end{eqnarray}
which is obviously periodic on the ring. Comparing Eqns (\ref{Isiper}) and (%
\ref{m-pt}), it is manifest how closely the correlations of the
non-equilibrium model mirror those of the Ising chain. Again, the only
significant difference is the appearance of the two amplitudes, $A_{e}$ and $%
A_{o}$.

%

\end{document}